\documentclass[%
 reprint,
superscriptaddress,
 amsmath,amssymb,
 aps,
]{revtex4-2}

\usepackage{graphicx}% Include figure files
\usepackage{upgreek}
\usepackage{subcaption}
\usepackage{dcolumn}% Align table columns on decimal point
\usepackage{bm}% bold math
\usepackage{xcolor}

\begin{document}

\preprint{APS/123-QED}

\title{Chiral Vector Beams for Topological Enantio-Sensitive Imaging}

% Force line breaks with \\

%Use only initial letters for first name, no full stops; use superscript number\textmd{$^{1,2\dots}$} for affiliation (10 pt)  
%and symbol\textmd{$^{*,\dagger}$} for corresponding author(s) (2 max) in footnote (10 pt) %(maximum 2 emails)  
%end \authors

\author{A. Rodriguez}
\email{aude.rodriguez@icmm.csic.es}
\affiliation{Instituto de Ciencia de Materiales de Madrid (ICMM-CSIC), 28049 Madrid, Spain }
\author{J. Edginton}
\affiliation{Department of Chemistry, Molecular Sciences Research Hub, Imperial College London, SW7 2AZ London, UK}
\author{A. Löhr}
\affiliation{Max-Born-Institut, 12489 Berlin, Germany}
\author{J. Terentjevas}
\affiliation{Max-Born-Institut, 12489 Berlin, Germany}
\author{P. Vindel-Zandbergen}
\affiliation{Department of Chemistry, New York University, New York 10003, USA}
\author{A. Ordóñez}
\affiliation{Department of Physics, Freie Universität Berlin, 14195 Berlin}
\author{J. Vogwell}
\affiliation{Department of Physics, Imperial College London, SW7 2AZ London, UK}
\author{O. Smirnova}
\affiliation{Max-Born-Institut, 12489 Berlin, Germany}
\affiliation{Technische Universität Berlin, 10623 Berlin, Germany}
\affiliation{Technion - Israeli Institute of Technology, Haifa, 32000, Israel}
\author{D. Ayuso}
\email{d.ayuso@imperial.ac.uk}
\affiliation{Department of Chemistry, Molecular Sciences Research Hub, Imperial College London, SW7 2AZ London, UK}
\affiliation{Max-Born-Institut, 12489 Berlin, Germany}
\author{L. Rego}
\email{laura.rego@csic.es}
\affiliation{Instituto de Ciencia de Materiales de Madrid (ICMM-CSIC), 28049 Madrid, Spain }

\date{\today}

\begin{abstract}

Molecular chiral discrimination is vital across chemistry and life sciences, yet optical approaches remain highly inefficient for small organic molecules. Here we turn vector beams, whose polarization carries a topological charge, into an efficient probe of molecular handedness. We show that the combination of a tightly focused radially polarized beam at the fundamental frequency $\omega$ and an azimuthally polarized beam at $2\omega$ creates chiral vector beams, a locally chiral field with structured handedness. Using state-of-the-art computational modeling in the chiral molecule carvone, we show that the nonlinear response (notably sum-frequency generation) of randomly oriented chiral molecules to this chiral vector beam captures the fields' local handedness and is highly enantio-sensitive. The two enantiomers emit light at frequency $3\omega$ with topological ring structures of different divergence, and thus the molecular handedness can be read out directly from the ring radius. The signal inherits the Poincaré index of the driving field, opening up new ways to robust, ultrafast, and and all-optical chiral discrimination.

\end{abstract}

%Use showkeys class option if keyword Keywords: chiral imaging, vector beam, topological light, non-linear interactions
\maketitle

\begin{figure*}[t!]
    \centering
    \includegraphics[width=1\linewidth]{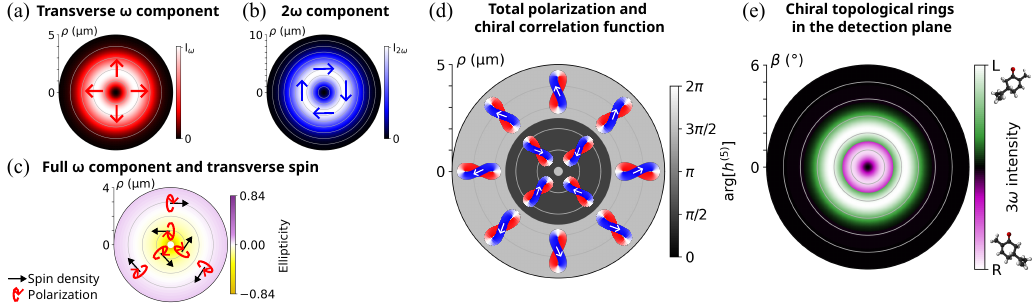}
    \caption{(a) Radially polarized vector beam at frequency $\omega$ and (b) azimuthally polarized vector beam at $2\omega$. The tight focus of (a) results in (c) an ellipticity in the $(\rho, z)$ plane (a mask has been applied to select the intensity above 8\%). This ellipticity ($E_z/E_\rho$) induces an azimuthally varying spin (black arrows), opposite in the inner and outer parts of the beam. These three components creates a (d) chiral vector beam, i.e. topological and azimuthally varying synthetic chiral light, quantified by the chiral correlation function $h^{(5)}$, whose phase dictates the chirality of the Lissajous figures. Red (blue) indicates a negative (positive) longitudinal component of the electric field. The delay $\psi_{2\omega}=0$ has yet not been optimized. (e) Observable chiral topological rings with different radius for different enantiomers.}
    \label{fig:fig1}
\end{figure*}

\begin{figure}[ht!]
    \centering
    \includegraphics[width=1\linewidth]{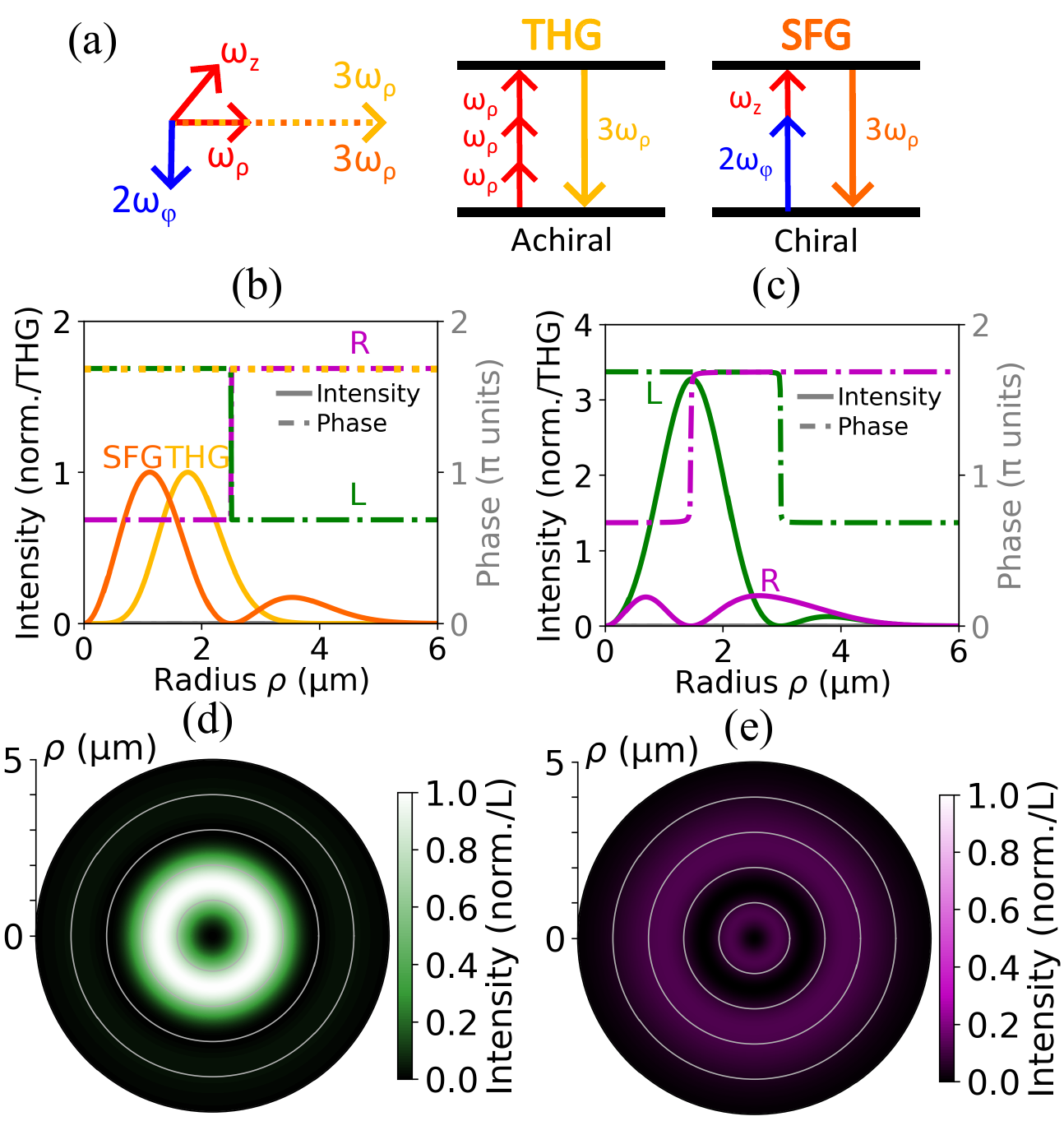}
    \caption{\textbf{Non-linear interactions in the near-field.} (a) Non-linear interactions between the topological beam and the chiral molecules, at each point of the interaction plane: THG (achiral) and SFG (chiral), both with radial polarization. (b,~c) Intensity (solid lines) and phase (dash-dotted lines) profiles of (b) the two non-linear contributions, SFG (orange) and THG (yellow), and (c) their sum for each enantiomer, L (green) and R (magenta), along the radial coordinate. (d,~e) Complete intensity profiles, resulting from the interference of THG and SFG, emitted by (d) left-handed and (e) right-handed enantiomers. Both profiles are normalized with respect to the left-handed enantiomer's emission.}
    \label{fig:fig2}
\end{figure}

Chirality is present at every scale of the universe: it describes the way galaxies rotate \cite{Galaxies}, impacts how snails mate \cite{snails_Maderspacher2016}, and constitutes a fundamental property of subatomic particles \cite{Griffiths1987}. An object is chiral if it is not superimposable with its mirror image, like a glove that would only fit one hand. The left- and right-handed versions of a chiral molecule (enantiomers) behave exactly the same, unless they interact with another chiral object. Since most biological molecules are chiral, enantiomeric discrimination is key in metabolic reactions and plays a vital role in drug development \cite{Francotte2006, Brooks2011}.

Chiroptical spectroscopy \cite{Berova2013, Forbes2026} has long relied on weak linear effects arising beyond the electric-dipole approximation \cite{Barron2004} and suffers from weak enantio-sensitivity.
Indeed, the differential absorption of circularly polarized light from left and right handed molecules is typically below $0.1\%$.
Novel approaches relying on purely electric-dipole interactions enable orders-of-magnitude greater sensitivity via detection of asymmetric photoelectron currents~\cite{PECD_Bowering2001, PECD_Lux2012, janssen2014, Wanie2024}, molecular fragments \cite{Fehre2021, Pitzer2013, Milner2019}, or electromagnetic radiation (i.e. all-optical setups).

The possibility of shaping light's polarization in 3D ~\cite{SCL_Ayuso2019, Ayuso2021NatComm, Opportunities_Ayuso2022, khokhlova2022, SCL_Rego2023, Ye2023, topoChirality_Mayer2024, Smirnova2025, ordonez2025, Gefei2026, Greenwood2026, Neufeld2019, Greenwood2025, Kohnke2026, tilting_Rodriguez2026} and making it locally chiral~\cite{SCL_Ayuso2019, Ayuso2021NatComm, Opportunities_Ayuso2022, khokhlova2022, SCL_Rego2023, Ye2023, topoChirality_Mayer2024, Smirnova2025, ordonez2025, Gefei2026} opens up exciting opportunities for chiral sensing and manipulation~\cite{ordonez2025} in the nonlinear regime of purely electric-dipole interactions. Two recent experiments \cite{Gefei2026,Greenwood2026} recording highly enantio-sensitive signals from isotropic chiral media establish this synthetic chiral light as a powerful and versatile platform for nonlinear chiral optics.
In addition, by combining spin and orbital angular momentum (OAM) in vortex beams, one can endow locally chiral light with topological properties to drive optical signals that are highly enantio-sensitive as well as robust with respect to experimental imperfections and noise~\cite{topoChirality_Mayer2024}.

Here we show how, using vector beams~\cite{VectorBeam_Zhan2009, Forbes2026}, which do not carry net OAM but instead carry a topological charge associated to their spatially varying polarization (Poincaré index), we can create a new type of structured light for efficient chiral recognition in thin samples: chiral vector beams.
Using state-of-the-art computational modeling, we show that the nonlinear response of randomly oriented chiral molecules to such chiral vector beams captures the molecular handedness, while inheriting the driving beam's topology and axial symmetry. This leads to spatial separation of the radiation emitted at new optical frequencies.

Chiral vector beams can be created by combining two vector beams of different frequencies. Consider the superposition of tightly focused radial and azimuthal beams carrying $\omega$ and $2\omega$ frequencies, respectively, as illustrated in Figs. \ref{fig:fig1}(a,b). They both carry a Poincaré index of $P=1$, defined as the variation of the polarization angle in a closed loop around the singularity, divided by 2$\pi$.
The steep intensity gradient along the radial $\omega$-field's polarization produces a longitudinal component that is stronger than that of a focused Gaussian beam~\cite{FocusedVB_Dorn2003}. This component exhibits a strong on-axis field, phase-delayed by $\pi/2$ with respect to the radial field, and a weak external ring with opposite phase (see Supplemental Material). Thus, the $\omega$ field acquires transverse spin.
Crucially, the direction of the transverse spin's vector $\vec{S}_{\omega} \propto \text{Im}(\vec{E}_\omega^*\times\vec{E}_\omega)$ captures the sign of the longitudinal component and is fundamentally locked to light's propagation (spin-momentum locking).
As a result, $\vec{S}_{\omega}$ flips direction at the $\omega$-beam waist radius $\rho=w_\omega$, see Fig.~\ref{fig:fig1}(c).
Note that, because the polarization of the $2\omega$ field is orthogonal to the intensity gradient, see Fig.~1(b), $\vec{S}_{2\omega}=\vec{0}$.
The superposition of the radial $\omega$ and azimuthal $2\omega$ field realizes a locally chiral field: the electric-field vector traces a 3D chiral structure in one optical cycle of the $\omega$ frequency at each point in space, see Fig.~\ref{fig:fig1}(d).
The field's chirality can be characterized via the 5th order chiral correlation function~\cite{SCL_Ayuso2019} $h^{(5)} \propto -(\vec{E}_{\omega})^2\, \vec{S}_{\omega}\cdot\vec{E}_{2\omega}$, which records the relative orientation between the $\omega$ spin and $2\omega$ polarization vectors. 
That is, the topology of $\vec{S}_{\omega}$ is imprinted onto the spatial distribution of handedness.
Consequently, the field has opposite chirality in the inner and outer parts of the ring-shaped spatial profile, thus realizing two chiral rings of opposite handedness.
The spatial distribution of the field's handedness determines the enantio-sensitive response of randomly oriented chiral molecules, leading to separation of their nonlinear optical emission, as shown in Fig. \ref{fig:fig1}(e).

We have modeled the low-order nonlinear response of a 2D sample of randomly oriented carvone molecules to our topologically structured chiral field by solving the time-dependent Schr\"odinger equation, taking into account macroscopic effects, see Supplemental Material.
We focus on the enantio-sensitive response at frequency $3\omega$, resulting from the interference between sum-frequency generation (SFG) \cite{Fischer2000PRL, Fischer2002CPL, Belkin2001PRL, Moad2004, Fu2013, Yan2014, Wang2015, Perets2022} from the $\omega$ and $2\omega$ fields, and third-harmonic generation (THG) from the $\omega$ field~\cite{SFG_Vogwell2023}, see Fig.~\ref{fig:fig2}(a).
Such an interference is enantio-sensitive because SFG, which is exclusive of chiral media, has opposite phase in opposite molecular enantiomers (chiral emission), whereas THG is not sensitive to the medium's handedness (achiral emission).
We considered laser wavelengths of $\lambda_{\omega} = 639$~nm and $\lambda_{2\omega}  = 320$~nm (thus $2\omega$ is close to the energy of the first excited state of the carvone molecule), pulse durations of 50~fs, intensities of $I_\omega = 2.4\times 10^{8}$~W/cm$^2$ and $I_{2\omega} = 2.4\times10^6$~W/cm$^2$, phase delay of $\psi_{2\omega} = 1.94\pi$, and beam waists of $w_{\omega} =2.5~\upmu$m and $w_{2\omega}= 5~\upmu$m, which are within reach of current experimental capabilities~\cite{FocusedVB_Dorn2003}. For these parameters, SFG and THG are emitted with a wavelength of $\lambda_{3\omega}=213$~nm and
$I_{\omega}$ was adjusted to optimize the enantio-sensitive interference.
Other nonlinear processes giving rise to $3\omega$ emission are negligible for these laser parameters. For a thin 3-$\upmu$m liquid jet, we estimate the maximum intensity of the nonlinear response in the near field of the order of 10$^{-4}$~W/cm$^2$ (see Supplemental Material).

Figure~\ref{fig:fig2}(b) shows the intensity of THG (yellow) and SFG (orange) in the near field as a function of the radial coordinate $\rho$. The cylindrical symmetry of the vector beam is conserved during the nonlinear interaction, and thus the SFG and THG fields are also cylindrically symmetric.
Likewise, both fields maintain the on-axis singularity at $\rho=0$.
However, whereas the THG response exhibits a single-ring shape, the SFG response has a double-ring structure, vanishing where the longitudinal $\omega$-field component is zero, i.e. where $\vec{S}_\omega=\vec{0}$ [Fig. \ref{fig:fig1}(b)].
This occurs because the two multiphoton processes involve different field components:  $\vec{P}_{\text{SFG}}= \varepsilon_0 \chi^{(2)}\,\vec{E}_{\omega}\times \vec{E}_{2\omega}$ and $\vec{P}_{\text{THG}}=\varepsilon_0 \chi^{(3)}\,(\vec{E}_{\omega}\cdot\vec{E}_{\omega})\,\vec{E}_{\omega}$, where $\vec{P}_i$ are the induced polarization densities, $\varepsilon_0$ is the vacuum permittivity and $\chi^{(2)}$ and $\chi^{(3)}$ are the corresponding second- and third-order susceptibilities (being $\chi^{(2)}_R=-\chi^{(2)}_L$ and $\chi^{(3)}_R=\chi^{(3)}_L$). Note that both SFG and THG contributions in the transversal plane have radial polarization.

The change of phase of $\pi$ in the longitudinal $\omega$-field component at its radial singularity is imprinted onto the chiral SFG response.
As a result, one can adjust the $2\omega$ phase delay in the driving field so that, for the L-enantiomer, the SFG-THG interference is constructive in the inner part of the ring and destructive in the outer part, leading to a more intense response at frequency $3\omega$, as shown by the green curves of Fig.~\ref{fig:fig2}(c) and the 2D intensity profile in Fig.~\ref{fig:fig2}(d).
As the SFG response has opposite phase in opposite enantiomers, we find the opposite situation in the R-enantiomer, whose response is more suppressed, leading to two concentric rings of similar amplitude, see magenta curves in Fig. \ref{fig:fig2}(c) and Fig. \ref{fig:fig2}(e).
A $\pi$-shift in the two-color phase delay of the driving field is equivalent to exchanging the molecular enantiomer, as it reverses the sign of the SFG emission. The radial nodes of Figs.~\ref{fig:fig2}(c-e) appear when SFG and THG reach the same amplitude at the same radius, and THG's phase is opposite to SFG's.

\begin{figure}[!ht]
    \centering
    \includegraphics[width=1\linewidth]{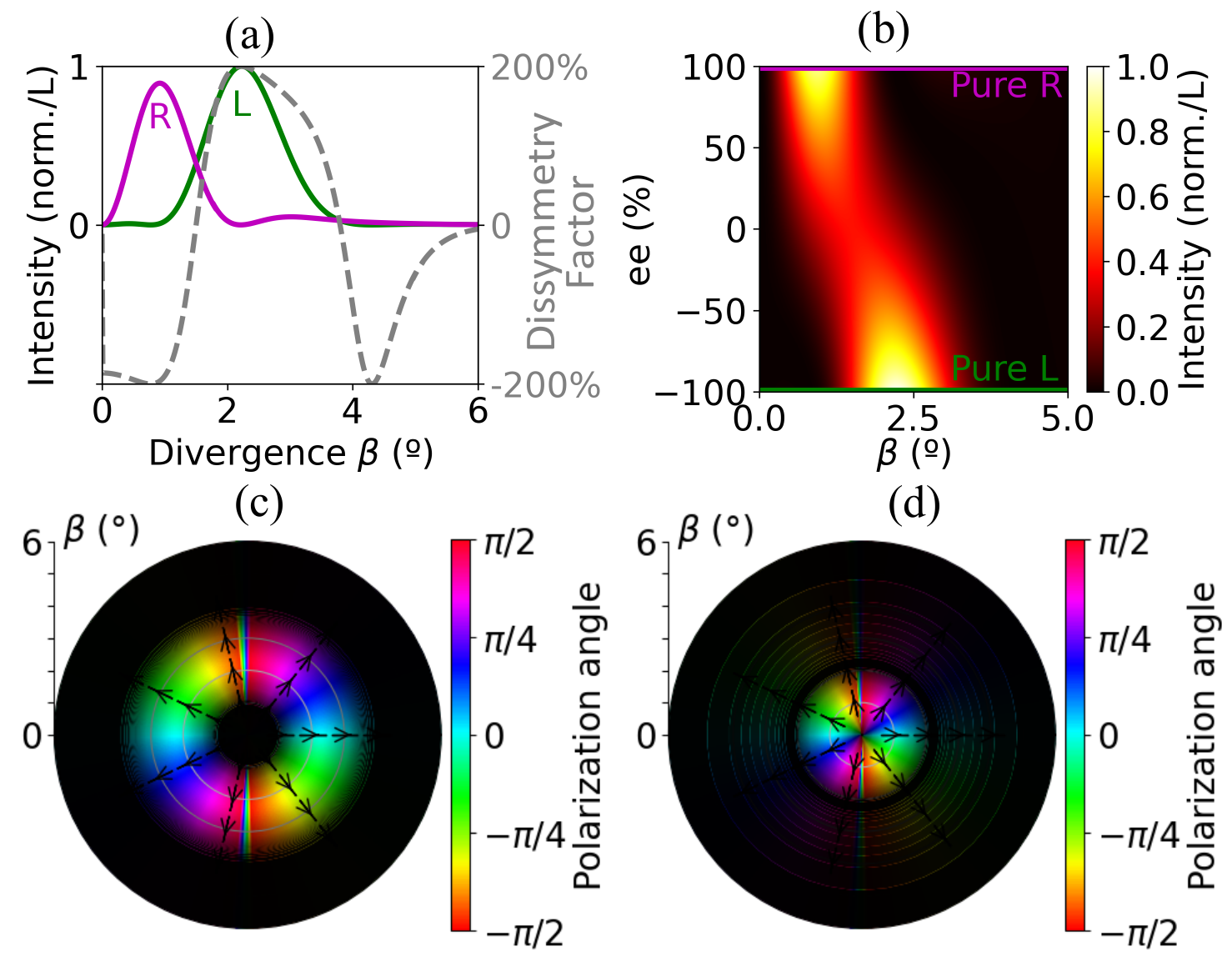}
    \caption{%\textbf{Enantio-sensitive far-field intensity profiles.} The total intensity after propagating to the far field emitted from (a) left-handed and (b) right-handed enantiomers. Both profiles are normalized with respect to the right-handed enantiomer's emission. (c) Intensity (solid lines) profiles for each enantiomer, and dissymmetry factor in gray (dashed line). (d) Displacement of the intensity's maximum (i.e. the ring) for a mixture of different enantiomeric proportions, quantified by the enantiomeric excess.
    %New version:
    \textbf{Enantio-sensitive far-field intensity profiles.} (a) Radial variation of the intensity (solid lines) profiles for each enantiomer, and dissymmetry factor (gray dashed line). (b) Displacement of the intensity's maximum (and thus the ring's divergence) for a mixture of different enantiomeric proportions, quantified by the enantiomeric excess $ee$. Polarization angle of the emissions from the (c) L- and (d) R-enantiomers, where the transparency is modulated according to the local intensity, normalized with respect to the L-enantiomer.
    }
    \label{fig:fig3}
\end{figure}

After propagation to the detection plane, the emission presents enantio-sensitive far-field intensity profiles, as shown in Fig.~\ref{fig:fig3}. Indeed, the molecular handedness is encoded in the radius of the main ring, with a difference of more than $1^\circ$ in divergence between the two enantiomers; therefore, an iris can be used to spatially filter out the exterior ring (i.e. the contribution of one of the enantiomers).
The degree of enantio-sensitivity can be quantified by the dissymmetry factor $g=2(I_L-I_R)/(I_L+I_R)$, where $I_L$ and $I_R$ are the intensity emitted by the L- and R-enantiomers, respectively. As shown in Fig.~\ref{fig:fig3}(a), our calculations reveal that $g$ reaches the $200\%$ limit at each radial node of either emissions (i.e. where one of the enantiomer's emission is suppressed). The emitted beam's divergence allows as well the determination of the enantiomeric excess in mixtures, $ee=(C_R-C_L)/(C_R+C_L)$, where $C_L$ and $C_R$ are the concentration of left- and right-handed molecules. This is shown in Figure~\ref{fig:fig3}(b), that represents the radial variation of the coherent addition of L- and R-enantiomers' emissions as a function of the $ee$.
The 2D far-field profiles for different $ee$ values are shown in the Supplemental Material.

The topology of the driving beam is transferred to the nonlinear emission, which preserves the same Poincaré index as the chiral vector beam ($P=1$). This is proved by the spatially varying polarization angle, displayed in Figs.~\ref{fig:fig3}(c,d), and calculated from the Stokes parameters, for the L- and R- enantiomers. The $3\omega$ emissions, that are radially polarized in the near field, keep their radial polarization after propagation, with the same winding number and axial symmetry. The transfer of the driving beam's Poincaré index to the nonlinear emission implies that the latter acquires the distinct properties of vector beams, such as enhanced robustness upon propagation~\cite{Nape2022, Cox2020, Zhu2021}.

\begin{figure}[ht!]
    \centering
    \includegraphics[width=1\linewidth]{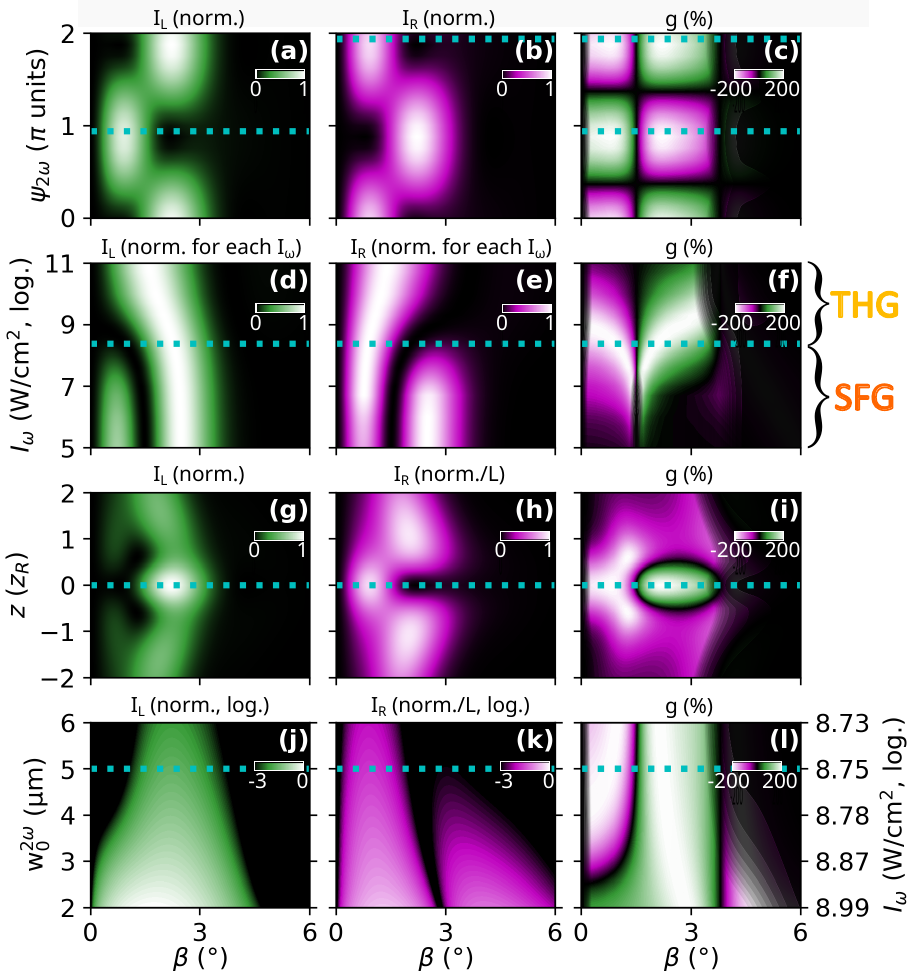}
    \caption{\textbf{Impact of the laser field parameters on the far-field enantio-sensitivity.} (a-c) Influence of the delay $\psi_{2\omega}$ on the intensity profiles of the (a) L- and (b) R-enantiomers and (c) the dissymmetry factor. (d-f) Influence of the intensity of the $\omega$ beam, $I_{\omega}$, on the intensity profiles of the (d) L- and (e) R-enantiomers and (f) the dissymmetry factor. The SFG and THG labels indicate the dominant phenomena. (g-i) Influence of the longitudinal distance, $z$, between the focus and the target on the intensity profiles of the (g) L- and (h) R-enantiomers and (i) the dissymmetry factor. The $z$ coordinate is shown in units of  the Rayleigh range of the $\omega$ beam ($z_R=30\ \upmu$m). (j-l) Influence of the beam waist $w_{2\omega}$ on the intensity profiles of the (j) L- and (k) R-enantiomers and (l) the dissymmetry factor. For each waist value, $I_\omega$ has been optimized to equalize the amplitude of the SFG and THG effects in the near field. In all panels, the relation $I_{2\omega} = 0.01 I_{\omega}$ is maintained, and the dotted blue lines indicate the values with the best enantio-sensitivity for each parameter. A  transparency mask has been added to the dissymmetry factor and only the regions where the intensity is above 10\% of the maximum are shown.}
    \label{fig:fig4}
\end{figure}

As our method relies on nonlinear interactions, it provides several control knobs to harness the enantio-sensitive response through the laser parameters (Fig.~\ref{fig:fig4}).
The low-order enantio-sensitive signal considered here results from the interference between SFG and THG, and therefore both responses must be balanced and phase locked.
Since only the SFG pathway involves $2\omega$ absorption, the SFG--THG interference can be controlled through the $2\omega$ phase delay, which in turn enables enantio-sensitive control over the divergence angle of the nonlinear emission, see Figs.~\ref{fig:fig4}(a-c). This delay can also be used for calibration in order to apply the method to any chiral molecule with an unknown response: one must find the delay for which the nonlinear emission differ the most spatially and the $I_{2\omega}/I_{\omega}$ ratio that maximizes such double-ring shape. Indeed, as SFG and THG are nonlinear processes of different order, involving the absorption of two vs three photons, changes in the beams' intensities alter their relative strength, enabling maximal interference, as shown in Figs.~\ref{fig:fig4}(d-f). 

Figs.~\ref{fig:fig4}(g-i) show the influence on the intensity profiles of a longitudinal misalignment of the beams focus with respect to the target position. As we chose a bigger waist for the $2\omega$-beam, it is less affected by an eventual misalignment (its Rayleigh range is 245~$\upmu$m). The displacement of the $\omega$ beam affects both the radial and longitudinal components at the target position and therefore both THG and SFG emissions. This exchanges the dominant enantiomer from $0.5z_R$ on. Note that optimizing the intensity and the delay would change the shape and interaction of SFG and THG and therefore allow to compensate the longitudinal misalignment. Therefore, Fig.~\ref{fig:fig4} also demonstrates that experimental fluctuations leading to small deviations in the beam's phase-delay, intensity, focus position or waist still allow the retrieval of large dissymmetry factors.

Another interesting control knob is the waist of the azimuthal beam $w_{2\omega}$ [Figs.~\ref{fig:fig4}(j-l)]. Indeed, for fixed $w_{\omega}$, the $2\omega$-beam controls the enhancement of the interior ($\rho<w_{\omega}$) or the exterior part ($\rho>w_{\omega}$) of the SFG signal. Thus, a small $w_{2\omega}$ ($<4$~$\upmu$m) allows the suppression of one enantiomer (here, the R-enantiomer). A higher $w_{2\omega}$ results in weaker emissions, but spatially separated chiral rings, obtained for the first time in the present work.

In this letter, we propose a topological imaging technique that uses chiral vector beams for efficient chiral discrimination, reaching maximum values of dissymmetry factor (200\%).
Here, the nonlinear emission of each enantiomer forms chiral rings with different radii.  
Indeed, we observe an enantio-sensitive divergence, paving the route for the experimental distinction of enantiomers, for instance, by inserting an iris to suppress the emission from a chosen molecular handedness. Which of the two enantiomers produces the larger beam radius is kept in our control through the delay between both beams. This delay, together with $I_{2 \omega}$, can be used to maximize the enantio-sensitivity for any chiral molecule, thanks to the linear scaling of the phase and amplitude of the SFG signal with the delay and amplitude of the $2\omega$ field. Thus, one can enforce an enantiomer to show the same intensity profile as the green curve in Fig.~\ref{fig:fig3}(a), and the opposite enantiomer to exhibit the magenta profile. 

Furthermore, our results are stable with respect to experimental fluctuations, such as intensity deviations, misalignment of the focus or jitter. 
Secondly, due to the coherence of the light-matter interaction, the topological charge of the driving field is imprinted into the nonlinear emission, which form vector beams with the same Poincaré index, inheriting the intrinsic robustness of vector beams~\cite{Nape2022, Cox2020, Zhu2021}.
Thirdly, the collinear geometry provides an alternative to the challenges of non-collinear configurations (beam alignment, delay control, angle-dependent detection, spatial/temporal walk off, etc \cite{SFGandVB_Ji2024, Garling2019}) and we exploit low-order nonlinear optical interactions that require low laser intensities and are readily integrated into ultrafast pump–probe experiments without vacuum infrastructure.
Therefore, our topological configuration allows for highly enantio-sensitive, ultrafast, all optical enantio-discrimination.

\section*{\label{sec:ack} Acknowledgments}
The authors acknowledge fruitful discussions with Drs. Margarita Khokhlova, Emilio Pisanty and Nicola Mayer. A. R. and L. R. acknowledge that the project leading to these results have received funding from “la Caixa” Foundation (ID 100010434), under the agreement “LCF/BQ/PR24/12050018”, and the Spanish Ministry of Science, Innovation and Universities \& the State Research Agency through the project ref. PID2024-163024NA-I00 (MICIU/AEI/10.13039/501100011033/FEDER, UE), and support from the Severo Ochoa Centers of Excellence program through Grant CEX2024-001445-S. 
P.~V.~Z. is grateful to the Simons Foundation and NYU IT High Performance Computing for the computational resources. 
A. O. acknowledges funding from the Deutsche Forschungsgemeinschaft (DFG, German Research Foundation) 543760364. J.~E. and D. A. aknowledge Funding from Royal Society URF$\backslash$R$\backslash$251036.
A. L., J. T. and O. S. acknowledge funding/co-funding by the European Union (ERC, ULISSES, 101054696). Views and opinions expressed are however those of the author(s) only and do not necessarily reflect those of the European Union or the European Research Council. Neither the European Union nor the granting authority can be held responsible for them.

%\nocite{*}

\bibliography{SFG/SFGbib.bib}% Produces the bibliography via BibTeX.

\setcounter{section}{0}
\setcounter{figure}{0}
\makeatletter
\renewcommand{\thefigure}{S\@arabic{\c@figure}}
\makeatother
\setcounter{figure}{0}
\renewcommand{\theequation}{S\arabic{equation}}
\preprint{APS/123-QED}

\title{Supplemental material for: Using Vector Beams for Topological Chiral Imaging \\with Sum-frequency Generation}% Force line breaks with \\

\author{Aude Rodriguez}
\email{aude.rodriguez@icmm.csic.es}
\affiliation{
 Instituto de Ciencia de Materiales de Madrid (ICMM-CSIC), 28049 Madrid, Spain
}%
\author{Josh Vogwell}
\affiliation{
 Department of Physics, Imperial College London, SW7 2AZ London, UK
}%
\author{David Ayuso}
\affiliation{
 Department of Chemistry, Molecular Sciences Research Hub, Imperial College London, SW7 2AZ London, UK
}%
\author{Laura Rego}
\affiliation{%
 Instituto de Ciencia de Materiales de Madrid (ICMM-CSIC), 28049 Madrid, Spain
}%

\date{\today}

\section*{Supplemental Material} 
\maketitle

%\tableofcontents

\section{\label{sec:1av}Microscopic description: single-molecule response}

To compute the ultrafast interaction between the laser field and the carvone molecules, the time dependent Schrödinger equation (TDSE) was solved using dipoles generated from state of the art ab-initio density functional theory (DFT) methods. 
In particular, the molecular geometry of the most stable R-carvone conformer was taken from \cite{lambertopticalactivitycarvone2012} and reoptimized using DFT \cite{kohndensityfunctionaltheory1996} with the B3LYP \cite{leedevelopmentcollesalvetticorrelationenergy1988,beckedensityfunctionalexchangeenergyapproximation1988} functional and the 6-311G(d,p) \cite{mcleancontractedgaussianbasis2008} basis set available in Gaussian 16 \cite{ g16}. Time-dependent DFT with CAM-B3LYP/d-aug-cc-pVDZ \cite{YANAI200451, dunning89, augccpvdz} was used to evaluate the excitation energies and transition dipoles for the first 100 excited states. The transition dipoles between the excited states were computed with the Multiwfn software \cite{multiwfn2012}.
The laser field considered in our TDSE simulations is a combination of two colors with orthogonal polarization:
\begin{equation}\textbf{E}_0(t) = a(t)[E_0^{\omega} \cos(\omega_0 t)\hat{\textbf{e}}_x+E_0^{2\omega} \cos(2\omega_0 t)\hat{\textbf{e}}_z],
\label{eq:laser_field_TDSE}
\end{equation}
where $\omega_0$= 0.07148 a.u. is the fundamental frequency, $E_0^{\omega}=6.81\times 10^{-5}$~a.u. and $E_0^{2\omega}=3.14\times 10^{-7}$~a.u. are the peak electric field amplitudes used in the microscopic calculations, and $a(t)$ is a Gaussian envelope centered at time zero with a full width half maximum of 50 fs. The relative strength of the third-harmonic generation and sum-frequency generation was calculated by separately computing averaged microscopic dipoles in the presence of only the fundamental field and in the presence of both colors, respectively. Note that the intensities described here are the ones used in the TDSE calculations, while the ones in the main text are the ones used for the scaling after optimization.

From the TDSE calculations, we compute the microscopic time-dependent dipoles for 442 orientations of the carvone molecules. The corresponding average induced polarization is:
$$P(t)=\frac{1}{8\pi^2} \int_0^{2\pi} \int_0^{\pi} \int_0^{2\pi} P^{\phi, \theta, \chi}(t) \sin(\theta)\
d\phi\ d\theta\ d\chi,
$$
where $P^{\phi, \theta, \delta}$ is the polarization induced by molecules with an orientation of $\phi$, $\theta$ and $\delta$. We integrate over $\phi$ and $\theta$ with a Lebedev quadrature of order 7, and over $\delta$ using the trapezoid method with 17 equally spaced grid points.

Under this configuration, Four-Wave Mixing (FWM) ($2\omega$+2$\omega$-$\omega$), which is a competing achiral interaction, can be neglected in this molecule (below $0.5\%$ of the THG signal), as long as $I_{\omega} \geq 100 I_{2\omega}$. Indeed, even if the FWM is present, it only contributes to the achiral part, and the present results still remain valid. Similar results, adjusting the delay, the waist and the intensity, can be obtained for other chiral molecules. 

After performing a Fourier transform to the calculated averaged induced polarization, we select the value of the spectrum at harmonic 3, obtaining the induced polarization for THG, $P_0^{\text{THG}}$, and SFG, $P_0^{\text{SFG}}$.

\section{\label{sec:2vb}Vector beam}
The driving beam is a superposition of two vector beams~\cite{VectorBeam_Zhan2009}: a radially polarized vector beam, $E_\rho$, of frequency $\omega$ and an azimuthally polarized vector beam, $E_\phi$, of frequency $2\omega$ with a delay of $\psi_{2\omega}$. Both are defined as a superposition of two circularly polarized vortex beams of opposite helicity and topological charge $\ell$ (associated to orbital angular momentum) and no radial modes $p$:
\begin{eqnarray}
    \begin{cases}
    \textbf{E}_{\rho}(\rho, \phi, z) =& \text{LG}_{\ell=1,p=0}^{\omega}(\rho, \phi, z)(\hat{\textbf{e}}_x+i\hat{\textbf{e}}_y)\\
    &+ \text{LG}_{\ell=-1,p=0}^{\omega}(\rho, \phi, z)(\hat{\textbf{e}}_x-i\hat{\textbf{e}}_y)\\
    \textbf{E}_{\phi}(\rho, \phi, z) = 
    &[\text{LG}_{\ell=1,p=0}^{2\omega}(\rho, \phi, z)(\hat{\textbf{e}}_x+i\hat{\textbf{e}}_y) \\
   & - \text{LG}_{\ell=-1,p=0}^{2\omega}(\rho, \phi, z)(\hat{\textbf{e}}_x-i\hat{\textbf{e}}_y)]e^{i\psi_{2\omega}},
    \end{cases}
\end{eqnarray}
where the subscript indicates the polarization direction.

The radial and azimuthal polarizations can be explicitly obtained by considering the polarization direction in polar coordinates ($\hat{\textbf{e}}_\rho=\cos \phi \hat{\textbf{e}}_x - \sin \phi\hat{\textbf{e}}_y $, $\hat{\textbf{e}}_\phi =\sin \phi \hat{\textbf{e}}_x + \cos \phi\hat{\textbf{e}}_y$), so the driving field can be rewritten as:

\begin{equation}
    \begin{cases}
    \textbf{E}_{\rho}(\rho, \phi, z) = 2R_{\ell=1,p=0}^{\omega}(\rho,z)\hat{\textbf{e}}_\rho\\
    
    \textbf{E}_{\phi}(\rho, \phi, z) = 2R_{\ell=1,p=0}^{2\omega}(\rho,z)  e^{i\psi_{2\omega}}\hat{\textbf{e}}_\phi,\\
    \end{cases}
    \label{eq:vb_rho_phi}
\end{equation}
with $R(\rho,z)$ the $\phi$-independent part of the Laguerre-Gaussian mode (defined in~\cite{LGmodes_Les1992}) at the focus point, so that $\text{LG}_{\ell=\pm 1,p}(\rho, \phi, z) = R_{\ell,p}(\rho,z)e^{ \pm i\phi}$. In the following, we will consider $z=0$ (the chiral molecules are placed at the focal plane of the beam):

\begin{equation}
    R_{\ell=1,p=0}^{i} (\rho) = A_{i}
    \frac{\sqrt{2}\rho}{w_i}
    e^{-\frac{\rho^2}{w_i^2}},
    \label{eq:R_lp}
\end{equation}
where $i=\{\omega, 2\omega\}$, $A_i$ is a constant, proportional to the maximum amplitude of the field, and $w_i$ is the beam waist. 
Note that the waist $w_i$ of a vector beam has a different physical meaning than that of a gaussian beam. For Laguerre-Gauss beams, the beam's size, which corresponds to the radius where the intensity drops below $1/e^2$, depends on $\ell$ and $p$ as $\sigma = w_i \sqrt{2p+|\ell|+1}$~\cite{phillips1983}. Thus, for the vector beams considered in our work (composed of Laguerre-Gauss modes with $\ell=\pm 1$), the beam's size is $\sigma=\sqrt{2} w_i$.

Upon tight focusing, a longitudinal component appears~\cite{Maxwell_Lax1975} at frequency $\omega$ (the azimuthally polarized beam does not exhibit a longitudinal component):
\begin{equation}
    \textbf{E}_{z}(\rho)  = \frac{i}{k}  \frac{1}{\rho} \frac{\partial }{\partial \rho}[\rho E_{\rho}(\rho)] \hat{\textbf{e}}_z.
\end{equation}

\begin{figure}[ht!]
    \centering
    \includegraphics[width=1\linewidth]{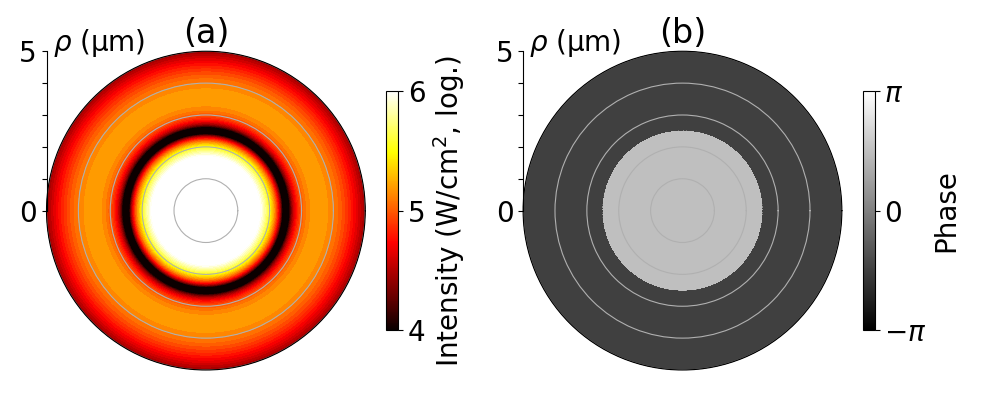}
    \caption{Transverse (a) intensity and (b) phase profiles of the longitudinal component. 
    }
    \label{fig:E_z}
\end{figure}

For $z=0$, $\ell=1$, $p=0$:
\begin{equation}
    {\textbf{E}}_{z}(\rho)= \frac{2i}{k} \frac{\sqrt{2}}{w_\omega} 
    A_0 
    \left(2-\frac{2\rho^2}{w^2_\omega}\right)
    e^{-\frac{\rho^2}{w^2_\omega}} \hat{\textbf{e}}_z.
    \label{eq:vb_z}
\end{equation}
Thus, $\textbf{E}_{z}(\rho)$ has a radial node at $\rho=w_\omega$, as pictured in Fig.~\ref{fig:E_z} ($\pi$-shift of the phase).

For $z\neq0$, the longitudinal component becomes:
\begin{align}
    {\textbf{E}}_{z}(\rho)=  i\frac{2\sqrt{2}}{k} \frac{w_\omega}{w(z)} 
    A_0 
    \left[2-
    \left(
    \frac{2}{w^2(z)}
    +i\frac{k}{R(z)}
    \right)\rho^2\right] \nonumber \\
    \exp\left({-\frac{\rho^2}{w^2(z)}}\right)
    \exp\left({i\frac{k}{2R(z)}\rho^2 -i\arctan\left(\frac{z}{z_R}\right) }\right) 
    \hat{\textbf{e}}_z,
    \label{eq:vb_z_zdep}
\end{align}

with $z_R=\pi\frac{w_\omega^2}{\lambda_\omega}$ the Rayleigh length of the $\omega$ beam, $w(z)=w_\omega\sqrt{1+(z/z_R)^2}$ the $z$-dependent waist, and $R(z)=z[1+(z_R/z)^2]$ the radius of curvature.

The Poincar\'e index of the beam can be computed as: 
\begin{equation}
    P=\frac{1}{2 \pi} \oint_C d \theta,
    \label{eq:poincare}
\end{equation}
where $\theta$ is the angle of the electric field vector with respect to the x axis, which is derived from its definition as a winding number \cite{Fulton1995}.

\section{Chiral Correlation Function}

The nonzero modulus of $h^{(5)}$ shown in Fig.~\ref{fig:fig1}(e) proves that the field is locally chiral (its polarization draws a 3D Lissajous figure in time) and its phase shows the spatial variation of the field's handedness. For our particular case, the origin of our chiral correlation function can be derived from the induced nonlinear polarization at frequency $3\omega$, which includes both SFG and THG:
\begin{align}
   | \textbf{P}(3\omega) |^2 &=  \big| \chi^{(2)}  \textbf{F}(2\omega) \times \textbf{F}(\omega) + \chi^{(3)} [ \textbf{F}(\omega) \cdot \textbf{F}(\omega)] \textbf{F}(\omega)  \big|^2  \nonumber \\ 
   & =  \text{(diagonal terms)} + \chi^{(2)} \chi^{(3)*} h^{(5)} +c.c.,
    \label{eq:P3w}
\end{align}
where $\chi^{(2)}$ and $\chi^{(3)}$ are the second- and third-order nonlinear susceptibilities associated to SFG and THG, respectively, and $h^{(5)}$ is:
\begin{align}
    h^{(5)}(& 2\omega, \omega, -\omega, -\omega, -\omega) \nonumber \\
    & \equiv 
    [\textbf{F}(2\omega)\times \textbf{F}(\omega)] [\textbf{F}^*(\omega)\cdot \textbf{F}^*(\omega)] \cdot \textbf{F}^*(\omega) \nonumber \\
    & = \textbf{F}(2\omega) \cdot[\textbf{F}(\omega)\times \textbf{F*}(\omega) ] [\textbf{F*}(\omega)\cdot \textbf{F*}(\omega)] \\
    & = \textbf{F}(2\omega) \cdot \frac{1}{2}[\textbf{F}(\omega)\times \textbf{F*}(\omega)+\textbf{F}(\omega)\times \textbf{F*}(\omega)] \nonumber \\
    & [\textbf{F*}(\omega)\cdot \textbf{F*}(\omega)] \nonumber \\
     & = \textbf{F}(2\omega) \cdot \frac{1}{2}[\textbf{F}(\omega)\times \textbf{F*}(\omega)-\textbf{F*}(\omega)\times \textbf{F}(\omega)]  \nonumber \\
     & [\textbf{F*}(\omega)\cdot \textbf{F*}(\omega)] \nonumber \\
     & = \textbf{F}(2\omega) \cdot \text{Im}[\textbf{F}(\omega)\times \textbf{F*}(\omega)] [\textbf{F*}(\omega)\cdot \textbf{F*}(\omega)] \nonumber \\
     &=-2\omega\textbf{F}(2\omega) \cdot \textbf{S}(\omega) [\textbf{F*}(\omega)\cdot \textbf{F*}(\omega)].
    \label{eq:h5_SFGTHG}
\end{align}
Note that the off-diagonal terms in equation (\ref{eq:P3w}) correspond to the interference between SFG and THG, which leads to the enantio-sensitive response. While the nonlinear susceptibilities characterize the material's response, $h^{(5)}$ only depends on the electric field.
Our time-dependent field, $\textbf{F}$, is defined as: 
\begin{equation}
    \textbf{F}(t)= (\underbrace{\textbf{E}_\rho + \textbf{E}_z)}_{\textbf{F}(\omega)}e^{-i\omega t} + \underbrace{\textbf{E}_\phi }_{\textbf{F}(2\omega)}e^{-2i\omega t} + \text{c.c.},
\end{equation}
where $\textbf{E}_\rho=E_\rho\textbf{e}_\rho$ and $\textbf{E}_\phi=E_\phi e^{i\psi_{2\omega}}\textbf{e}_\phi$ are given by~(\ref{eq:vb_rho_phi}) and $\textbf{E}_z=iE_z \textbf{e}_z$ by~(\ref{eq:vb_z}). 

Then, equation~(\ref{eq:h5_SFGTHG}) becomes:
\begin{align}
    h^{(5)}( &2\omega, \omega,-\omega,-\omega,-\omega)  \nonumber \\
     & = \textbf{F}(2\omega) \cdot[\textbf{F}(\omega)\times \textbf{F*}(\omega) ] [\textbf{F*}(\omega)\cdot \textbf{F*}(\omega)] \nonumber \\
     & = \textbf{E}_\phi\cdot[(\textbf{E}_\rho+\textbf{E}_z)\times (\textbf{E}_\rho-\textbf{E}_z)] \nonumber\\
     & [(\textbf{E}_\rho-\textbf{E}_z)\cdot (\textbf{E}_\rho-\textbf{E}_z)] \nonumber\\
     & =2iE_\rho E_\phi
     E_z(E_\rho^2-E_z^2)e^{i \psi_{2\omega}}.  
\end{align}
Note that the same expression can be derived from equation~(16) in \cite{SCL_Ayuso2019}.

On the other hand, $h^{(5)}$ can be factorized to obtain the Poincaré index of the nonlinear emission by expressing it as a product of vectors whose direction coincides with the polarization of the SFG and THG emissions, respectively:
\begin{align}
  \textbf{V}_{SFG} &= \textbf{F}(2\omega) \times \textbf{F}(\omega)=|\textbf{V}_{SFG}| \,\textbf{e}_{\rho}  \\ 
   \textbf{V}_{THG} &= [ \textbf{F}(\omega) \cdot \textbf{F}(\omega)] \textbf{F}(\omega) =|\textbf{V}_{THG}| \, \textbf{e}_{\rho} .
\end{align}
The Poincar\'e indices of SFG and THG can be written as 
\begin{align}
  P^{\text{SFG}}&=\frac{1}{2\pi} \oint_C d\theta_{SFG}=  \frac{1}{2\pi}\oint_C d\phi =1  \\
  P^{\text{THG}}&= \frac{1}{2\pi} \oint_C d\theta_{THG}=\frac{1}{2\pi}\oint_C d\phi =1  .
\end{align}
where $\theta_i=\text{arccos}(\textbf{V}_i \cdot \textbf{e}_x/|\textbf{V}_i|)$ and, given that both SFG and THG are radially polarized, $\theta_i=\text{arccos}(\textbf{e}_{\rho} \cdot \textbf{e}_x)=\phi$.
If $P^{\text{SFG}}=P^{\text{THG}}$, then $P^{\text{SFG}}=P^{\text{THG}}=P^{3 \omega}$ and the $3\omega$ emission conserves the same enantio-sensitivity for all azimuths, which ensures that structure of the emission is ring-shaped. This is equivalent to ensuring that $h^{(5)}$ is constant along the azimuthal coordinate (note that $h^{(5)} =\textbf{V}_{SFG} \cdot \textbf{V*}_{SFG}$). 

\section{\label{sec:NFsup}Macroscopic response: the near field}

The interaction between the described driving field and the chiral molecules results in two interfering non-linear phenomena SFG and THG, polarized in the radial direction. We model the spatial variation of the microscopic response from the TDSE calculations by considering a perturbative description. That means that we scale the microscopic response ($P^{\text{THG}}_0$ and $P^{\text{SFG}}_0$ from section I) with the beam from section II (eqs. \ref{eq:vb_rho_phi} and \ref{eq:vb_z}). On the one hand, the THG results from the absorption of three $\omega$ photons:
\begin{equation}    P^{\text{THG}}_{\rho}(\rho) = P^{\text{THG}}_0\left(\frac{E_\rho(\rho)}{E^{\omega}_{0}}\right)^3.
\end{equation}

On the other hand, SFG results from the absorption of one $2\omega$ photon and one $\omega$ photon. It presents a $\pi$ phase shift between left-handed and right-handed enantiomers:
\begin{equation}    P^{\text{SFG,R}}_{\rho}(\rho) = P^{\text{SFG}}_0\left(\frac{E_\phi(\rho)}{E^{2\omega}_{0}}\right)\left(\frac{E_z(\rho)}{E^{\omega}_{0}}\right) = - P^{\text{SFG,L}}_{\rho}(\rho).
\end{equation}

The total nonlinear induced polarization in the near field $P^{\text{NF}}_{\rho}$ is the addition of both responses:
\begin{equation}
    P^{\text{NF}}_{\rho}(\rho) = P^{\text{THG}}_{\rho}(\rho) + P^{\text{SFG, L/R}}_{\rho} (\rho),
\end{equation}
which is enantio-sensitive and it is polarized in the radial direction: $\textbf{P}^{\text{NF}}_{\rho}(\rho)=P^{\text{NF}}_{\rho}(\rho) \hat{\textbf{e}}_{\rho}$. The spatially-dependent induced polarization at a fixed frequency is proportional to the corresponding radiated power via the Larmor formula.

\section{\label{sec:FFsup}Propagation to the far field} 
\begin{figure}[ht!]
    \centering
    \includegraphics[width=0.8\linewidth]{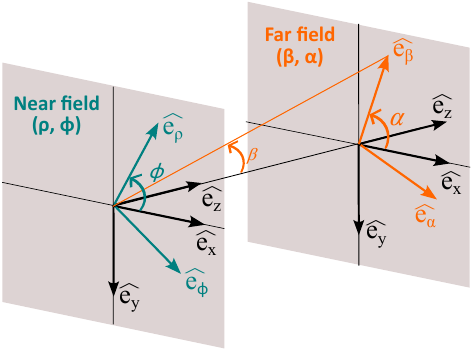}
    \caption{Definition of the spatial coordinates in the near and far field.}
    \label{fig:S1_coordinates}
\end{figure}

To obtain the nonlinear emission at the detection plane (far field), we apply the Fraunhofer diffraction equation, valid for $z_{\text{FF}} >\!>  \frac{\rho^2}{\lambda_{3\omega}}\approx 2\text{~mm}$, where $z_{\text{FF}}$ is the position of the detector with respect to the focus point. The radial and azimuthal far-field coordinates are denoted as $\alpha$ and  $\beta$, respectively (see Fig.~\ref{fig:S1_coordinates}). The two field polarization components in the far-field are denoted as $P_{\alpha}^{\text{FF}} (\alpha, \beta)$ and $P_{ \beta}^{\text{FF}}(\alpha, \beta)$ and they are obtained as:
\begin{eqnarray}
   & P_{\alpha, \beta}^{\text{FF}}&(\alpha, \beta)  \propto \\
 && \int\limits^{\infty}_0 \int\limits^{2\pi}_0 
    A_{\alpha, \beta}^{\text{NF}}(\alpha, \rho, \phi)
    e^{-i\frac{2\pi}{\lambda_{3\omega}} \tan\beta \cos(\alpha-\phi) \rho}\rho\ 
    d\phi\ d\rho \nonumber
\end{eqnarray}
where 
\begin{equation}
    \begin{cases}
    A_{\beta}^{\text{NF}}(\alpha, \rho, \phi) = \cos(\phi-\alpha)P_{\rho}^{\text{NF}}(\rho)\\
    
    A_{\alpha}^{\text{NF}}(\alpha, \rho, \phi) = \sin(\alpha-\phi)P_{\rho}^{\text{NF}}(\rho).\\
    \end{cases}
\end{equation}
In our configuration, $\lambda_{3\omega}=\lambda_{\omega}/3=213$~nm is the wavelength of the considered third harmonic. Note that the contribution of the longitudinal nonlinear polarization is neglected as it is suppressed due to phase matching.

\section{Intensity profiles for different concentrations}

\begin{figure}[!ht]
    \centering
    \includegraphics[width=\linewidth]{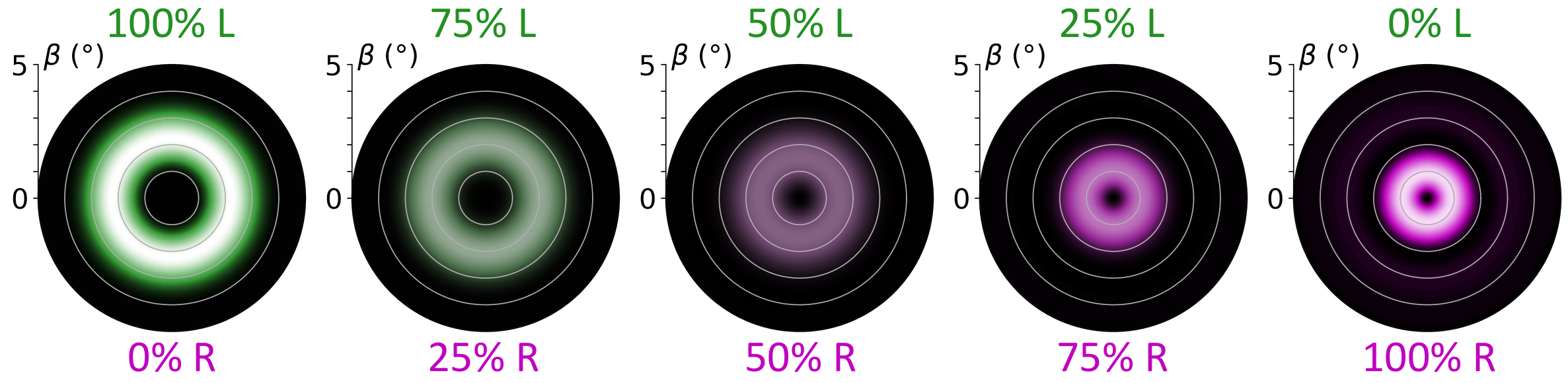}
    \caption{Far field intensity profiles of the enantio-sensitive SFG+THG for different concentrations of each enantiomer in a mixture.}
    \label{fig:S_diff_concentration}
\end{figure}

In figure \ref{fig:S_diff_concentration}, we show how changing the concentration of left-handed or right-handed molecules in a mixture affects the intensity profiles of the enantio-sensitive SFG in the far field. The radius of the main ring allows the identification of the proportion of each enantiomer, as shown in Fig.~\ref{fig:fig3}(b) in the main text.

\section{Estimation of the magnitude of the nonlinear emission in a thin liquid jet}

We estimate the order of magnitude of the nonlinear emission in a $3 \ \upmu$m liquid jet of carvone molecules.

The unidirectional pulse propagation equation (UPPE) derived from the Maxwell equations connects the induced polarization in the medium with the driving electric field. When written in the spectral domain and for the nonlinear emission at $3\omega$, the UPPE takes the following form: 
\begin{eqnarray}
      \partial_z E_{3\omega}(z)=i k_{3\omega} E_{3\omega} (z) 
      +i\frac{(3\omega)^2}{2\epsilon_0 c^2 k_{3\omega}}P_{3\omega} (z),  \label{eq:prop}
\end{eqnarray}
where ionization effects have been neglected. The solution of this propagation equation would require solving the TDSE at each value of the propagation distance $z$, which would become very computationally demanding. As our goal is to obtain an estimation of the order of magnitude of the signal, we neglect all propagation effects on the driving beams and approximate the induced polarization to be constant along $z$, expect for its $z$-dependent phase: $P_{3\omega}(z) \approx P_{3\omega}(z=0)e^{ik_{3\omega}z}$. For the initial condition of $E_{3\omega}(z=0)$, the solution of Equation \ref{eq:prop} would provide an approximate estimation of the intensity of the nonlinear emission at the end of the jet:
\begin{eqnarray}
      I_{3\omega}(L)\approx \frac{L^2 (3\omega)^2}{8 \epsilon_0 c}|P_{3 \omega}|^2,  \label{eq:esti_mono}
\end{eqnarray}
where $L=3\ \upmu$m is the thickness of the jet and $P_{3\omega}=\rho_{molec} d_{3\omega}$, $\rho_{molec}$ being the liquid's molecular number density and $d_{3\omega}$ the dipole obtained from the TDSE calculations after orientational averaging. Note however that this expression considers a monochromatic field, while we are using short laser pulses. To account for this, we introduce a temporal-averaged intensity of the dipole:
\begin{eqnarray}
      |d_{3 \omega}|^2 \rightarrow \frac{\int |d_{3 \omega} (t)|^2 dt}{\int dt}=\frac{\int |d_{3 \omega} (\omega')|^2 d\omega'}{2\pi T},  \label{eq:dipole_multichromatic}
\end{eqnarray}
where $T$ is the pulse duration and the integral over $\omega'$ is performed between $\omega'=2.5 \omega$ and  $\omega'=3.5 \omega$ to account for the spectral bandwidth of the nonlinear emission. Thus, Equation \ref{eq:esti_mono} takes the following form:
\begin{eqnarray}
      I_{3\omega}(L)\approx \frac{L^2  \rho_{molec}^2 (3\omega)^2}{8 \epsilon_0 c}\frac{\int |d_{3 \omega} (\omega')|^2 d\omega'}{2\pi T}.
\end{eqnarray}

\end{document}